\documentstyle[preprint,aps,12pt]{revtex}
\begin{document}
\draft
\hfill\vbox{\baselineskip14pt
            \hbox{\bf ETL-xx-xxx}
            \hbox{ETL Preprint 00-xxx}
            \hbox{April 2000}}
\baselineskip20pt
\vskip 0.2cm 
\begin{center}
{\Large\bf Quantum group conjecture and Stripes}
\end{center} 
\vskip 0.2cm 
\begin{center}
\large Sher~Alam
\end{center}
\begin{center}
{\it Physical Science Division, ETL, Tsukuba, Ibaraki 305, Japan}
\end{center}
\vskip 0.2cm 
\begin{center} 
\large Abstract
\end{center}
\begin{center}
\begin{minipage}{14cm}
\baselineskip=18pt
\noindent
Keeping in mind our conjecture of modelling cuprates and related systems 
by quantum groups and our claim that the existence of stripes is intimately 
related to quantum groups and more strongly is a consequence of the underlying
quantum group structure we note that a recent claim that {\em magnetic 
fluctuations} also displays a one-dimensional character makes our conjecture 
more viable. The recent strong claim by Mook et al., if true 
further supports our intuition. For the fluctuations associated with a striped
phase are expected to be one-dimensional, whereas the magnetic fluctuations 
are {\em taken} to display two-dimensional symmetry. It is claimed by Mook 
et al. that this apparent two-dimensionality results from measurements on 
twinned crystals, and that similar measurements on substantial detwinned 
crystals of YBa$_{2}$Cu$_{3}$O$_{6.6}$ reveal the one-dimensional character of 
the magnetic fluctuations thus making the stripe scenario more stronger, in 
other words there are also magnetic stripes. As noted before in our conjecture
that one of the main reason for quantum groups to model cuprates is that they 
are directly related to the symmetries of one-dimensional systems. Thus now if
the magnetic fluctuations are also shown exhibiting one-dimensional behavior 
that all the more {\em directly strengthens our conjecture}. However we note 
that our conjecture is more generic and not tied to the one-dimensional 
character of the now claimed magnetic fluctuations. Simply the one-dimesional 
nature of magnetic fluctuations as noted by Mook et al., makes our conjecture
more {\em obvious} and clearer. 


\end{minipage}
\end{center}
\vfill
\baselineskip=20pt
\normalsize
\newpage
\setcounter{page}{2}
	In a previous work one of us \cite{alam98} have advanced 
the conjecture that one should attempt to model the phenomena of
antiferromagnetism and superconductivity by using quantum
symmetry group. Following this conjecture to model the phenomenona of
antiferromagnetism and superconductivity by quantum symmetry
groups, three toy models were proposed \cite{alam99-1}, namely,
one based on ${\rm SO_{q}(3)}$ the other two constructed with
the ${\rm SO_{q}(4)}$ and ${\rm SO_{q}(5)}$ quantum groups. 
Possible motivations and rationale for these choices are 
were outlined. In \cite{alam99-2} a model to describe quantum 
liquids in transition from 1d to 2d dimensional crossover using 
quantum groups was outlined. Recently we \cite{alam00} also proposed 
the particular choice of classical group SO(7) to incorporate
the phenomenon of pseudo-gap not addressed in the SO(5) model.     
In \cite{alam00-1} we considered an idea to construct a theory 
based on patching critical points so as to simulate the behavior 
of systems such as cuprates and specific system was cited
as an example.

	In this short note we point out the recent experimental
work by Mook et al. \cite{moo00} which claims the one dimensional
nature of magnetic fluctuations in the YBa$_{2}$Cu$_{3}$O$_{6.6}$
system and quantum group conjecture. Unlike other works where
striped phases have been suggested to account for the
properties of high T$_{c}$ copper oxide materials. In our
formulation the existence of stripes is intimately
connected with the quantum groups.

	Stripes may considered as inhomogeneous distributions
of charge and spin. Naively if charge and spin are confined
to separate linear regions we can say that the systems is
in a striped phase. The separation of spin and charge as
originally proposed by Anderson is linked to the formation
of the stripe phase. The separation of spin and charge
in turn is intimately related to quantum group. The
quantum group allows the system to be broken into 
1-d quantum liquids.  

	From the neutron scattering data one can see
the existence of both the spin and charge fluctuations.
These fluctuations of spin and charge can be taken to imply
the existence of a dynamic striped phase for 	
YBa$_{2}$Cu$_{3}$O$_{7-x}$ materials with high T$_{c}$
value. However it has been noted that for two of the
cuprates namely YBa$_{2}$Cu$_{3}$O$_{7-x}$ and 
La$_{2-x}$Sr$_{x}$Cu$_{3}$O$_{7-x}$ the magnetic fluctuations
arising from spin fluctuations seem to exhibit a four-fold
pattern around the incommensurate points around the
$(1/2,1/2)$ reciprocal lattice positions.
However Mook et al. \cite{moo00}, have recently claimed
that this two-dimensional symmetry exhibited by
magnetic fluctuations is a result of using {\em twinned}
crystals in the neutron scattering studies.
For if the crystal is substantially detwinned 
the magnetic fluctuations becomes {\em one-dimensional}.
This means that not only the charge but spin
fluctuations exhibit a one-dimensional character
which further directly strengthens the quantum group scenario.
Moreover as pointed out earlier\cite{alam98,alam99-1,alam99-2}
to get the complete picture of cuprates one may have
to go beyond even the quantum group picture. As we
patch the 1-d sub-systems which are a consequence of
the quantum group symmetry to obtain the 2-d and
3-d systems we arrive naturally at a string picture.
The sequence of a possible scenario is quantum group
symmetry yields the 1-d system, the quantum groups
are also related to Kac-Moody algebras which in
turn are the generators of special non-linear sigma
models, which are themselves an approximate form
of strings \cite{kak91,alam00,alam00-1}. 

	Considering this scenario we note that in  
approximation it suffices that one can considers the 
interactions of spinons and holons. However one may 
regard spinons and holons as modes of a string and so 
in a more general picture one must expand
the string [stripe] in term of the basic components
such as spinons and holons. It is then possible
to regard that as the strings [stripes] interact from the 
2-d point of view the quantum group symmetry is `broken'. 

	Most people have tended to regard it a problem 
that quantum groups were closely tied with 1-d systems. 
However we take the contrary point of view, namely it is
the close relationship between the 1-d systems
and quantum groups which is useful in understanding
the cuprates and related systems since the basic
fluctuations in these systems are 1-d.
	
	The interaction of charge fluctuations with
phonons is highly non-trivial. The charge scattering
as imaged by phonons was reported in \cite{moo98,moo00}
for  YBa$_{2}$Cu$_{3}$O$_{6.6}$ using twinned
crystals in a region of momentum space where
it is expected a strong interactions between phonons and
the striped phase. It is seen that the broad line shape
[due to charge fluctuations] can be considered to be made 
of two `lines' one from ${\bf b}$ twin domain [which
is a normal narrow line] and a broad line from 
${\bf a}$ twin domain that contains the striped phase
1-d modulation vector. This is consistent with the
experimental findings mentioned in \cite{alam99-1}
about charge stripes and the phonon anomaly.	  

	In conclusion experiments seem to indicate the
one-dimensional nature of both charge and magnetic
fluctuations. The one-dimensional nature of fluctuations
is a consequence of quantum group symmetry. Experimental
evidence suggests that the superconductivity in the
cuprates is strongly influenced by the one-dimensional
stripe phases. Thus quantum group symmetry must play
a role in the understanding HTSC phenomenon.  
 
\section*{Acknowledgments}
The Sher Alam's work is supported by the Japan Society for
for technology [JST]. 


\begin{references}
\bibitem{alam98}Sher~Alam, {\em A Conjecture for possible theory for the 
description of high temperature superconductivity and 
antiferromagnetism}. Proceedings of Quantum Phenomena in Advanced 
Materials at High Magnetic Fields, 4th International Symposium on Advanced
Physical Fields [APF-4]. KEK-TH-607, KEK Preprint 98-xxx, cond-mat/9812060.
\bibitem{alam99-1}Sher~Alam, {\em Quantum Group based Modelling for the 
description of high temperature superconductivity and
antiferromagnetism}.~KEK-TH-613, KEK Preprint 98-xxx, cond-mat/9903038.
\bibitem{alam99-2}Sher~Alam, ~{\em Theoretical modeling for quantum liquids 
from 1d to 2d dimensional crossover using quantum groups}.
KEK-TH-619. KEK Preprint 99-xxx, cond-mat/990345.
\bibitem{alam00}Sher~Alam et al., {\em The choice of the symmetry group 
for the cuprates} cond-mat/0004269.
\bibitem{alam00-1}Sher~Alam et al., ~{\em The patching of critical points 
using quantum group}, cond-mat/0004350.
\bibitem{moo00}H.~A.~Mook et al., ~{\em One-dimensional nature of the
magnetic fluctuations in YBa$_{2}$Cu$_{3}$O$_{6.6}$}, Nature,
Vol.~{\bf 404},~(2000),~ 730, cond-mat/0004362.
\bibitem{kak91}M.~Kaku, {\em Strings, Conformal Fields, and
Topology: An Introduction}, Springer-Verlag, 1991.
\bibitem{moo98}H.~A.~Mook and F.~Dogan, , Nature,
Vol.~{\bf 401},~(1999),~ 145.


\end{references}
\end{document}